\documentclass[prd,reprint,twocolumn,amssymb]{revtex4-1}
\usepackage{amsmath}
\usepackage{amssymb} 
\usepackage[cm]{fullpage}
\usepackage{graphicx} 
\usepackage{siunitx}
\usepackage{tabularx} 
\usepackage[colorlinks,linkcolor=blue,citecolor=blue,urlcolor=blue]{hyperref}
\usepackage[table,xcdraw]{xcolor}
\usepackage[utf8]{inputenc}

\newcommand{\beq}{\begin{equation}}
\newcommand{\eeq}{\end{equation}}

\newcommand{\rpar}[1]{\left(#1\right)}
\newcommand{\spar}[1]{\left[#1\right]}

\newcommand{\bd}{{\rm d}}

\newcommand{\sigmadot}{\dot{\sigma}}
\newcommand{\phidot}{\dot{\phi}}

\begin{document}

\title{The adiabatic/entropy decomposition in $P(\phi^I,X^{IJ})$ theories with multiple sound speeds} 

\author{Chris Longden}
\email{cjlongden1@sheffield.ac.uk}
\affiliation{Consortium for Fundamental Physics, School of Mathematics and Statistics, \\
		University of Sheffield, Hounsfield Road, Sheffield, S3 7RH, United Kingdom}

\date{\today}

\begin{abstract}
We consider $P(\phi^I,X^{IJ})$ theories of multi-field inflation and ask the question of how to define the adiabatic and entropy perturbations, widely used in calculating the curvature and isocurvature power spectra, in this general context. It is found that when the field perturbations propagate with different speeds, these adiabatic and entropy modes are not generally the fundamental (most natural to canonically quantise) degrees of freedom that propagate with a single speed. The alternative fields which do propagate with a single speed are found to be a rotation in field space of the adiabatic and entropy perturbations. We show how this affects the form of the horizon-crossing power spectrum, when there is not a single ``adiabatic sound speed'' sourcing the curvature perturbation. Special cases of our results are discussed, including $P(X)$ theories where the adiabatic and entropy perturbations are fundamental. We finally look at physical motivations for considering multi-speed models of inflation, particularly showing that disformal couplings can naturally lead to the kind of kinetic interactions which cause fields to have different sound speeds. 
\end{abstract}

\maketitle

\section{Introduction}

Due to its general success in generating a near-scale-invariant spectrum of primordial fluctuations, inflationary cosmology has become a widely-accepted component of our understanding of the history of the universe. The accumulation of data from Cosmic Microwave Background experiments such as WMAP \cite{Bennett:2012zja} and Planck \cite{Ade:2015lrj,Ade:2015tva} in recent years has only served to further vindicate the paradigm. Despite this, the nature of the mechanism behind inflation -- what kind of field or fields drive the expansion, and if gravity is still well described by general relativity in this epoch -- continues to elude us. A vast range of theoretical models with similar predictions have been proposed \cite{Martin:2013tda}, and with further experimental constraints on these models appearing to be a somewhat long-term prospect, theorists in the field are looking for new approaches and tests to complement and support the process of testing models of inflation. One such approach is the study of general actions that encompass a wide range of individual models in the literature, such as

\beq
S = \int \bd^4 x \sqrt{-g} \spar{\frac{1}{2} R + P(\phi^I,X^{JK})} \, , \ \ (I,J,K) = 1 \ldots N \, , \label{eq:generalPlagrangian}
\eeq

in which general relativity is coupled to an arbitrary Lagrangian $P$ which depends on $N$ fields $\phi^I$ and the kinetic terms $X^{JK} = - \partial_\mu \phi^J \partial^\mu \phi^K / 2$. This kind of action, studied in \cite{Arroja:2008yy,Langlois:2008qf}, covers an enormous range of models. Of course, the special cases of this action include several physically-motivated and extensively studied theories with scalar sectors possessing non-standard kinetic structure, such as Braneworld-motivated Dirac-Born-Infeld (DBI) \cite{Alishahiha:2004eh} fields and modified gravity theories in the Einstein Frame, such as Starobinsky inflation \cite{Starobinsky:1980te, vandeBruck:2015xpa}. In this class of models, the nonlinear dependence on kinetic terms means that field perturbations generally propagate at speeds not equal to unity. In particular, we note that it is possible, though not extensively studied in the literature \cite{Cai:2009hw,Pi:2011tv,vandeBruck:2015tna}, for each field to have a different propagation speed in principle. This may have interesting consequences in, for example, the calculation of non-Gaussianity \cite{vandeBruck:2016vlw}.

It is common when working with multiple-field theories to define so-called adiabatic ($\sigma$) and entropy ($s$) fields \cite{Gordon:2000hv,GrootNibbelink:2001qt,Wands:2002bn,DiMarco:2002eb,Byrnes:2006fr}, as the fields which uniquely source the curvature $(\mathcal{R})$ and isocurvature $(\mathcal{S})$ perturbations, respectively. One finds, however, that depending on the structure of kinetic terms in a Lagrangian, the mathematical expressions for these fields differ in each case. In this work, our primary goal is to understand how to construct these fields in the highly general class of models specified by eq. (\ref{eq:generalPlagrangian}), and what the implications of this are for studying inflationary models of this class. Calculations will be explicitly carried out for the $N=2$ case for the sake of simplicity and minimalism.

In section \ref{sec:construction} we will construct the adiabatic and entropy fields for a general two-field theory. Following this, in section \ref{sec:quantise} we will turn our attention to quantisation of fields and the calculation of the curvature power spectrum, sourced by the adiabatic field, particularly in cases with non-equal propagation speeds of the two fields where this is non-trivial. Significantly, we will show that the adiabatic and entropy perturbations can generally each depend on multiple sound speeds, and there is hence not always a single ``adiabatic sound speed'' sourcing the curvature power spectrum. Finally, in section \ref{sec:examples}, we will apply our results to some physically-motivated models and interesting subsets of the general $P(\phi^I,X^{IJ})$ Lagrangian to give examples of the applicability of our results, recovering some known results from the literature as special cases of our work along the way, before concluding in section \ref{sec:conclusion}.

\section{Construction of the adiabatic and entropy perturbations} \label{sec:construction}

The second order action in $P(\phi^I,X^{IJ})$ theories can be expressed in the form

\begin{align}
S_{(2)} = \frac{1}{2} \int \bd t \bd^3 x a^3  \Big[ K_{IJ} \dot{Q}^I \dot{Q}^J - \frac{1}{a^2} P_{<IJ>} \partial_i Q^I \partial^i Q^J & \nonumber \\  - N_{IJ} \dot{Q}^I Q^J - M_{IJ} Q^I Q^J  \Big] & \, , \label{eq:2oaction}
\end{align}

where $K_{IJ} = P_{<IJ>} + 2 P_{<MI><JK>} X^{JK}$ is the kinetic structure matrix, and $M_{IJ}$ and $N_{IJ}$ are mass and interaction terms whose particular forms can be found in the literature \cite{Langlois:2008qf}, but are not important for the discussion ahead. This is because we are interested in the sound speeds in such theories, which depend only $K_{IJ}$ and $P_{<IJ>}$. In these expressions, subscripts in angular parenthesis represent symmetrised derivatives with respect to kinetic terms, defined such that

\beq \label{eq:symderiv}
P_{<IJ>} = \frac{1}{2} \rpar{\frac{\partial P}{\partial X^{IJ}} + \frac{\partial P}{\partial X^{JI}}} \,.
\eeq

One can construct adiabatic and entropy perturbations $(Q^\sigma, Q^s)$ by finding the linear combination of field perturbations $Q^I$ which meets two conditions. Firstly, the gradient term in the second order action should be orthonormal in the adiabatic/entropy basis. That is, for a field redefinition $Q^I = e^I_{I'} Q^{I'}$, then the four free functions $e^I_{I'}$ should satisfy

\beq
P_{<IJ>} e^I_{I'} e^J_{J'} = \delta_{I'J'} \, . \label{eq:diagonalgradientcondition}
\eeq

However, because in general $P_{<IJ>} = P_{<JI>}$, we need only use three of our four degrees of freedom to satisfy this condition. We can hence set one of the $e^I_{I'}$ functions to zero without loss of generality at this stage. For concreteness, we choose $e^\phi_{\chi'} = 0$. For brevity of notation, we will also define $e^\phi_{\phi'} = A \, ,e^\chi_{\chi'} = B  \,,e^\chi_{\phi'} = -C$, so that our field redefinition looks like

\begin{align}
Q^\phi & = A Q^{\phi'} \, , \label{eq:Qphiredef} \\
Q^\chi & = B Q^{\chi'} - C Q^{\phi'} \label{eq:Qchiredef} \, .
\end{align}

%and the elements of the transformed gradient matrix $P_{I'J'} = P_{<IJ>} e^I_{I'} e^J_{J'}$ are,
%
%\begin{align}
%P_{\phi' \phi'} & = A^2 P_{\phi \phi} - 2 A C P_{\phi \chi} + C^2 P_{\chi \chi} \, , \\
%P_{\phi' \chi'} & = B \rpar{A P_{\phi \chi} - C P_{\chi \chi}} \, , \\
%P_{\chi' \chi'} & = B^2 P_{\chi \chi} \, .
%\end{align}

We then solve $ P_{<IJ>} e^I_{I'} e^J_{J'} = \delta_{I'J'}$ to find

\beq
A = \sqrt{\frac{P_{\chi \chi}}{|P|}} \, , \quad B = \frac{1}{\sqrt{P_{\chi \chi}}} \, , \quad  C = \frac{P_{\phi \chi}}{\sqrt{P_{\chi \chi} |P|}} \, .
\eeq

Secondly, our adiabatic  and entropy fields $Q^n \ (n = \sigma, s)$ should respectively be aligned parallel and perpendicular to the trajectory in ($\phi',\chi'$) field space. As we only used three of our four degrees of reparametrisation freedom so far, we are free to perform a one-parameter rotation of the fields to achieve this, that is, the adiabatic and entropy fields are constructed as

\begin{align}
Q^\sigma & = Q^{\phi'} \cos \theta + Q^{\chi'} \sin \theta \, , \label{eq:sigmatrig} \\
Q^s & = Q^{\phi'} \cos \theta - Q^{\chi'} \sin \theta \, , \label{eq:strig}
\end{align}

where $\cos \theta = \dot{\phi}' / \dot{\sigma}$, $\sin \theta = \dot{\chi}' / \dot{\sigma}$ and $\sigmadot^2 = \rho + P = P_{<IJ>} \phidot^I \phidot^J = P_{<I'J'>} \phidot^{I'} \phidot^{J'}$. As this only amounts to a rotation of the fields, the gradient term $\delta_{nm} = e^{I'}_n e^{J'}_m \delta{I' J'}$ remains orthonormal. Using eqs. (\ref{eq:Qphiredef}--\ref{eq:Qchiredef}), we can rewrite our expressions for $Q^n$ as

\begin{align}
Q^\sigma & = \frac{1}{A} \rpar{\frac{C}{B} \sin \theta + \cos \theta} Q^\phi + \frac{1}{B} \sin \theta \ Q^\chi \, , \label{eq:sigmatrig2} \\
Q^s & = \frac{1}{A} \rpar{ \frac{C}{B} \cos \theta - \sin \theta} Q^\phi + \frac{1}{B} \cos \theta \ Q^\chi  \label{eq:strig2} \, .
\end{align}

These expressions can then be further manipulated by using eqs. (\ref{eq:Qphiredef}--\ref{eq:Qchiredef}) to find  the time derivatives of the $\phi^{I'}$ fields in terms of the original fields

\begin{align}
\frac{\dot{\phi}'}{\sigmadot} & =  \cos \theta = \frac{1}{A \sigmadot} \dot{\phi} \, , \\
\frac{\dot{\chi}'}{\sigmadot} & = \sin \theta =  \frac{1}{B \sigmadot} \rpar{\dot{\chi} + \frac{C}{A} \dot{\phi}} \, .
\end{align}

Using this, it is easy to check for common simple cases like $P = X^{\phi \phi} + X^{\chi \chi}$ \cite{Gordon:2000hv} or $P = X^{\phi \phi} + e^{2b(\phi)} X^{\chi \chi}$ \cite{DiMarco:2002eb} that our results reduce to the usual definitions of the adiabatic and entropy fields. Putting the time derivatives of the fields together with eqs. (\ref{eq:sigmatrig2}--\ref{eq:strig2}), we finally obtain a construction of the adiabatic and entropy fields in terms of the original $\phi^I$ basis:

\begin{align}
Q^\sigma &= \frac{P_{<IJ>} \dot{\phi}^I Q^J}{\sigmadot} \, ,  \label{eq:Qsigmadef} \\
Q^s & = \frac{\sqrt{|P|}}{\sigmadot} \rpar{\dot{\chi} Q^\phi - \dot{\phi} Q^\chi}  \label{eq:Qsdef} \, .
\end{align}

Importantly, the adiabatic field (\ref{eq:Qsigmadef}) we have constructed satisfies

\beq
\mathcal{R} = \frac{H}{\sigmadot} Q^\sigma \, ,
\eeq

that is, it is the lone source of the curvature perturbation. 

\section{Quantisation and Sound Speeds} \label{sec:quantise}

In single-field inflation the second order action contains the terms

\beq
S_{(2)} \supset \frac{1}{2} \int \bd \eta \bd^3 x \Big[ (v')^2 - \frac{1}{a^2} c_s^2 \partial_i v\partial^i v + \ldots \Big]  \, ,
\eeq

such that the equation of motion in conformal time $\eta$, for a Fourier mode $k$ in the large-$k$ limit is

\beq \label{eq:singlefieldquant}
v_k'' - \frac{c_s^2 k^2}{a^2} v_k^2 = 0 \, ,
\eeq

where $c_s$ is the propagation speed of perturbations, given by the square root of the ratio of the gradient and kinetic terms in the action. Then, when performing canonical quantisation of the field to fix the boundary conditions of the solutions, one imposes that the vacuum asymptotically approaches the Minkowski vacuum for high frequency modes, that is,

\beq
v_k \rightarrow \frac{1}{2 c_s k} e^{-i c_s k \eta} \, . \label{eq:canonquant}
\eeq

In two-field theories, a complication in this process arises. In general, the equations of motions in the large $k$ limit will take the form

\begin{align}
 K_{\phi \phi} \ddot{Q}^\phi + K_{\phi \chi} \ddot{Q}^\chi  - \frac{k^2}{a^2} \rpar{ P_{<\phi \phi>} Q^\phi + P_{<\phi \chi>} Q^\chi} + \ldots & = 0 \, \label{eq:abc1} \\ 
 K_{\phi \chi} \ddot{Q}^\phi + K_{\chi \chi} \ddot{Q}^\chi  - \frac{k^2}{a^2} \rpar{ P_{<\phi \chi>} Q^\phi + P_{<\chi \chi>} Q^\chi} + \ldots & = 0 \, , \label{eq:abc2} 
\end{align}

where the omitted terms beyond the ellipses are those coming from the mass ($M_{IJ}$) and interaction ($N_{IJ}$) terms in the action (\ref{eq:2oaction}). As in the standard single field case of eq. (\ref{eq:singlefieldquant}), we quantise such that the vacuum is asymptotically Minkowski and thus assume these terms negligible at leading order. As there is a mixing of kinetic terms in $Q^\phi$ with gradient terms in $Q^\chi$ and so on, one cannot assign propagation speeds to these two variables -- they are not the fundamental propagating degrees of freedom. Instead, some linear combination of $Q^I$ fields with diagonalised gradient and kinetic matrices would be the canonically quantisable fields, and one would have to set initial conditions for fields like $Q^\phi$ as a linear combination of the initial conditions for the fundamental degrees of freedom. 

To construct the fields with diagonal kinetic and gradient matrices, we define another basis $Q^a = e^{I'}_a Q^{I'} ,\ a = (\psi, \omega)$, in which $K_{ab} = e^{I'}_a e^{J'}_b K_{I' J'}$ is to be a diagonal matrix. As the gradient term is already diagonal in the $Q^{I'}$ basis, the transformation $e^{I'}_a$ must be a rotation so as to not spoil this, in which case the components of $K_{ab}$ are

\begin{align}
K_{\psi\psi} & = K_{\phi' \phi'} \cos^2 \Theta - K_{\phi'\chi'} \sin 2 \Theta + K_{\chi' \chi'} \sin^2 \Theta  \, ,\\ \label{eq:Kpsipsi}
K_{\psi\omega} & = \frac{1}{2} \rpar{2 K_{\phi' \chi'} \cos 2 \Theta + (K_{\phi'\phi'} - K_{\chi' \chi'}) \sin 2 \Theta } \, ,\\
K_{\omega\omega} & = K_{\phi' \phi'} \sin^2 \Theta + K_{\phi'\chi'} \sin 2 \Theta + K_{\chi' \chi'} \cos^2 \Theta  \, . \label{eq:Komegaomega}
\end{align}

Inspecting the expression for the off-diagonal terms $K_{\psi\omega}$, it is clear that we need to rotate by an angle

\beq
\tan 2 \Theta = \frac{2 K_{\phi' \chi'}}{K_{\chi'\chi'} - K_{\phi' \phi'} } \, . \label{eq:abasisangle}
\eeq

\begin{figure}
    \centering
    \includegraphics[width=0.5\textwidth]{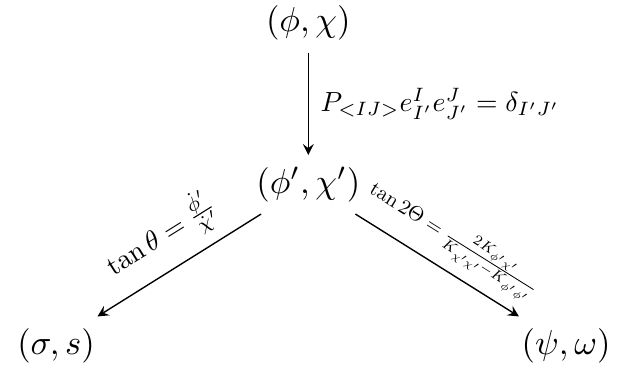}
    \caption{Relationships between different fields defined in this work. The $(\phi \chi)$ basis contains the original fields appearing in the action (\ref{eq:generalPlagrangian}). The $(\phi', \chi')$ basis is constructed to diagonalise the gradient term in the second-order action (\ref{eq:2oaction}) according to (\ref{eq:diagonalgradientcondition}). From this basis, we define two rotations in field space. The first of these is the entropy/adiabatic basis $(Q^\sigma, Q^s)$ in which the angle of rotation orients the fields parallel and perpendicular to the trajectory in field space (\ref{eq:sigmatrig}--\ref{eq:strig}). The second is the fundamental basis $(Q^\psi, Q^\omega)$ in which the angle of rotation is chosen such that the two fields are fundamental degrees of freedom propagating with pure, independent sound speeds.}
    \label{fig:fieldrelations}
\end{figure}

This angle is different to the angle $\theta$ which we rotated the $Q^{I'}$ fields by to obtain the adiabatic and entropy perturbations, and $(Q^\sigma, Q^s)$ are not in general the fundamental degrees of freedom. A summary of the different fields defined in this work and their relation to each other is shown as a diagram in figure \ref{fig:fieldrelations} for convenience.

In the ($Q^\psi, Q^\omega$) basis, the equations of motion (again, in the large $k$ limit, as in eqs. (\ref{eq:singlefieldquant},\ref{eq:abc1} -- \ref{eq:abc2})) then take the form

\begin{align}
\ddot{Q}^\psi  - \frac{(c_s^\psi)^2 k^2}{a^2} Q^\psi & = 0 \, \\
\ddot{Q}^\omega  - \frac{ (c_s^\omega)^2 k^2}{a^2}  Q^\omega & = 0 \, ,
\end{align}

where $(c_s^a)^2 = K_{aa}^{-1}$, and the $K_{aa}$ can be obtained by substituting eq. (\ref{eq:abasisangle}) into eqs. (\ref{eq:Kpsipsi}--\ref{eq:Komegaomega}) to find

\begin{align}
K_{\psi\psi} & =\frac{ K_{\chi' \chi'} + K_{\phi' \phi'} - \sqrt{\rpar{K_{\chi' \chi'} + K_{\phi' \phi'}}^2 - 4 |K_{I' J'}|}}{2}  \, ,\\
K_{\omega\omega} & = \frac{ K_{\chi' \chi'} + K_{\phi' \phi'} + \sqrt{\rpar{K_{\chi' \chi'} + K_{\phi' \phi'}}^2 - 4 |K_{I' J'}|}}{2}  \, ,
\end{align}

which are the eigenvalues of $K_{I' J'}$ as one would expect from the structure of the second order action, and of course, generally not the same. The adiabatic and entropy fields are then related to the ($\psi, \omega$) fields via a combination of the two rotations \footnote{Out of interest, we note the mathematical similarity in this procedure to that carried out in \cite{Byrnes:2006fr}, where a field redefinition that diagonalises the mass matrix instead of the kinetic structure (which is there assumed canonical).}, such that

\begin{align}
Q_\sigma & = \cos \rpar{\theta-\Theta} Q_\psi + \sin \rpar{\theta-\Theta} Q_\omega \, , \\
Q_s & =  \cos \rpar{\theta-\Theta} Q_\omega -  \sin \rpar{\theta-\Theta} Q_\psi \, , 
\end{align}

and it is from this, canonically quantising the $(\psi, \omega)$ fields in the spirit of (\ref{eq:canonquant}), we can obtain the form of the leading order power spectrum at horizon crossing when the two sound speeds are distinct:

\beq
\mathcal{P}_\mathcal{R}^* = \frac{H^2}{8 \pi^2 \epsilon} \spar{ \rpar{\frac{\cos \rpar{\theta-\Theta}}{\sqrt{c_s^\psi}}}^2 + \rpar{\frac{\sin \rpar{\theta-\Theta}}{\sqrt{c_s^\omega}}}^2} \, . \label{eq:multispeedspectrum}
\eeq

In the special case when the two sound speeds are equal ($c_s^\psi = c_s^\omega = c_s$) this reduces to the usual expression,

\beq
\mathcal{P}_\mathcal{R}^* = \frac{H^2}{8 \pi^2 \epsilon c_s} \, .
\eeq

This result is interesting as it shows that the leading order power spectrum only cares about the difference in angle between the fundamental fields and the adiabatic field when the two sound speeds of propagation are unequal. This makes sense because if the speeds of sound are equal then in the $(\psi, \omega)$ basis, the kinetic matrix $K_{ab}$ is proportional to the identity matrix, and as such any rotation of it (including the rotation into the adiabatic/entropy basis) will leave it unchanged, with no off-diagonal elements. Similarly, if $\theta = \Theta$ then the $(\sigma, s)$ basis and the $(\psi, \omega)$ basis are equivalent (the adiabatic/entropy fields are the fundamental degrees of freedom) and eq. (\ref{eq:multispeedspectrum}) reduces to 

\beq
\mathcal{P}_\mathcal{R}^* = \frac{H^2}{8 \pi^2 \epsilon c_s^\sigma} \, ,
\eeq

and while there may in general be a distinct second sound speed for entropy modes $c_s^s$, it does not affect the curvature perturbation. The condition for the adiabatic field being fundamental in this way can be found by comparing eq. (\ref{eq:abasisangle}) to eqs. (\ref{eq:sigmatrig}--\ref{eq:strig}), from which one can see that $\tan \theta = \dot{\phi}'/\dot{\chi}'$, to obtain the condition

\beq
\theta = \Theta \quad \implies \quad \frac{2 K_{\phi' \chi'}}{K_{\chi'\chi'} - K_{\phi' \phi'} } = \frac{2 \dot{\phi}'\dot{\chi}'}{\rpar{\dot{\chi}'}^2 - (\dot{\phi}')^2}  \, . \label{eq:psisigmacondition}
\eeq

Finally, there is also the special case where $\Theta = 0$, that is, the fundamental fields are $(\phi', \chi')$. From eq. (\ref{eq:abasisangle}) it is clear this occurs when $K_{\phi' \chi'} = 0$, which occurs for Lagrangians whose derivatives fulfill the condition $(P_{<\chi \chi>} P_{<M \phi > <\chi K>} - P_{<\phi \chi>} P_{< M \chi> < \chi K>} ) X^{MK}= 0$. We can write the power spectrum in this case as:

\beq
\mathcal{P}_\mathcal{R}^* = \frac{H^2}{8 \pi^2 \sigmadot^2 \epsilon}\spar{ \rpar{\frac{\phidot'}{\sqrt{c_s^{\phi'}}}}^2 + \rpar{\frac{\dot{\chi}'}{\sqrt{c_s^{\chi'}}}}^2} \, .
\eeq

\section{Examples} \label{sec:examples}

\subsection*{$P(X)$ theories}

A widely studied subclass of the Lagrangian studied here is the case when the kinetic part of the action is an arbitrary function of $X = G_{IJ} X^{IJ}$ \cite{Langlois:2008mn} where $G_{IJ}(\phi^K)$ is a field space metric. Performing the field redefinition (\ref{eq:Qphiredef}--\ref{eq:Qchiredef}), we obtain a kinetic matrix,

\beq
K_{I' J'} = \delta_{I' J'} + 2\frac{P_{XX}}{P_{X}} X_{I' J'} = \delta_{I' J'} + \frac{P_{XX}}{P_{X}} \dot{\phi}^{I'} \dot{\phi}^{J'} \, ,
\eeq

where in the second equality we used the fact that in the $(I', J')$ basis the field space metric is orthonormal, $X_{I' J'}$  has the same components as $X^{I' J'} = \dot{\phi}^{I'} \dot{\phi}^{J'} /2$. In such theories, we hence find that the condition (\ref{eq:psisigmacondition}) is fulfilled. That is, in this special case, the adiabatic and entropy modes are the fundamental degrees of freedom. We find that the components of the kinetic matrix are

\begin{align}
K_{\sigma\sigma} & = 1 + \frac{P_{XX}}{P_X} \rpar{X + \sqrt{X^2 - 4 |G||X|} }  \, ,\\
K_{ss} & = 1 + \frac{P_{XX}}{P_X} \rpar{X - \sqrt{X^2 - 4 |G||X|} }   \, .
\end{align}

Using $(X^{\phi\chi})^2 = X^{\phi \phi} X^{\chi \chi}$, or equivalently $|X| = X^{\phi \phi} X^{\chi \chi} - (X^{\phi\chi})^2 = 0$, this reduces to

\begin{align}
(c_s^\sigma)^2 &= \frac{P_X}{P_X + 2 P_{XX} X } \, , \\
(c_s^s)^2 & = 1  \, .
\end{align}

We hence see that when the function $P(X)$ is non-linear in $X$ such that $P_{XX} \neq 0$, this is a two-sound-speed model where only the adiabatic mode has a propagation speed differing from unity. A generalisation of this case where $P = P(Y)$ where $Y = X + b(X^2 - X^I_J X^J_I)$ was studied in \cite{Arroja:2008yy}, and it is found in this case that the adiabatic and entropy modes are still the fundamental degrees of freedom and the two sound speeds are still generally different, but the entropy sound speed need not be $1$. Instead, they find

\beq
(c_s^s)^2 = 1 + b X \, .
\eeq 

\subsection*{$P(X^{\phi \phi}, X^{\chi \chi})$ theories}

Considering theories where the Lagrangian is an arbitrary function of the single-field kinetic terms $X^{\phi \phi}$ and $X^{\chi \chi}$, but not of the explicit kinetic interaction $X^{\phi \chi}$, we find that the structure of the theory remains highly general as terms of the form $\alpha (X^{\phi \phi} X^{\chi \chi})^n$ are permitted, and at the background level $X^{\phi \phi} X^{\chi \chi} = (X^{\phi \chi})^2$. We find explicitly, in this case, that

\beq
K_{\phi' \chi'} = \frac{2 X^{\phi \chi} P_{<\phi \phi><\chi \chi>}}{\sqrt{P_{<\phi \phi>} P_{<\chi \chi>}}} \, ,
\eeq

which is nonzero when interaction terms containing factors of $X^{\phi \phi} X^{\chi \chi}$ are present in $P$. The angle $\Theta$ will then be correspondingly nonzero and the fundamental degrees of freedom $Q^\psi$ and $Q^\omega$ will propagate with non-trivial sound speeds. Kinetic interaction terms are hence still implicitly present and having an effect on the nature of perturbations in the theory unless we restrict ourselves further to the subclass in which

\beq
P = f(X^{\phi \phi}) + g(X^{\chi \chi}) \, .
\eeq

Here $P_{<\phi \phi><\chi \chi>} = 0$ and hence $\Theta = 0$ such that the $Q^{\phi'}$ and $Q^{\chi'}$ are the fundamental fields with sound speeds

\begin{align}
\rpar{c_s^{\phi'}}^2 &= \frac{f_{<\phi \phi>}}{f_{<\phi \phi>}+ 2f_{<\phi \phi><\phi \phi>} X^{\phi \phi} }  \, , \\
\rpar{c_s^{\chi'}}^2 &= \frac{g_{<\chi \chi>}}{g_{<\chi \chi>}+ 2g_{<\chi \chi><\chi \chi>} X^{\chi \chi} }  \,.
\end{align}

Note that these are non-equal even when $f = g$. Cases like this have been studied in e.g. \cite{Cai:2009hw,Pi:2011tv}.

\subsection*{Models from $\mathcal{N}=1$ Supergravity}

In $\mathcal{N}=1$ supergravity, the kinetic part of the scalar sector Lagrangian is given by \cite{Lyth:1998xn}

\beq
P = \frac{\partial^2 \mathcal{K}}{\partial \phi^I \partial (\phi^J)^*} \partial_\mu \phi^I \partial^\mu (\phi^J)^* \, ,
\eeq

where $\mathcal{K}$ is the K\"ahler potential. Irrespective whether the K\"ahler potential is minimal or nonminimal, this Lagrangian contains no nonlinearities in the kinetic terms ($P_{<MI><JK>} = 0$),  so $K_{IJ} = P_{<IJ>}$. This implies $K_{I' J'} = \delta_{I' J'}$ and hence from eq. (\ref{eq:abasisangle}), $\Theta = 0$ and the $Q^{I'}$ fields are the fundamental degrees of freedom. All of the fields will propagate with sound speeds of 1, as $K_{I',J'} = P_{I' J'} = \delta_{I' J'}$. 

\subsection*{Disformally coupled inflation}

Disformally coupled inflation \cite{vandeBruck:2015tna,vandeBruck:2016vlw} is a two-field model of inflation in which a scalar field ($\chi$) is confined to a brane whose induced metric $\hat{g}$ is disformally related to the metric of spacetime $g$. This is equivalent to a $P(\phi^I,X^{IJ})$ theory with kinetic structure

\beq
P = \frac{C}{D}\rpar{1 - \frac{1}{\gamma}} + \frac{C}{\gamma}X^{\chi \chi} + 2 \gamma D (X^{\phi \chi})^2 \, ,
\eeq

where $C$ and $D$ are conformal and disformal couplings relating the two metrics via $\hat{g}_{\mu \nu} = C g_{\mu \nu} + D \phi_{,\mu} \phi_{,\nu}$, and $\gamma^{-1} = \rpar{1 - 2 \frac{D}{C} X^{\phi \phi}}^{1/2}$. Despite the complicated kinetic structure containing interactions like $(X^{\phi \chi})^2$ and non-linearities coming from the factors of $\gamma$ in each term due to the presence of a kinetic term ($\phi_{,\mu} \phi_{,\nu}$) in the disformal metric, as well as the DBI-type kinetic term for $\phi$ (as it is related to the radial motion of the brane), it turns out that for this model, $K_{\phi' \chi'} = 0$ and so $\Theta = 0$ and the $(\phi', \chi')$ fields are the fundamental degrees of freedom, with sound speeds

\begin{align}
c_s^{\phi'}  = \sqrt{\frac{C - \gamma D p_\chi}{\gamma^2 C + \gamma D \rho_\chi}} \, , \quad 
c_s^{\chi'} = \frac{1}{\gamma} \, ,
\end{align}

where $p_\chi$ and $\rho_\chi$ are the pressure and energy density associated with the $\chi$ field. These sound speeds are non-equal when $D \neq 0$. We hence see that disformal couplings are a possible physical motivation for considering such multi-speed models.

\section{Conclusions} \label{sec:conclusion}

We constructed the adiabatic and entropy field perturbations in general $P(\phi^I,X^{IJ})$ models of inflation and, significantly, shown that they do not generally correspond to the fundamental, canonically quantised, degrees of freedom in scenarios when multiple distinct propagation speeds are present. While some existing work has computed adiabatic power spectra for special cases in which there are multiple sound speeds, the results we present here clarify how to correctly apply the adiabatic/entropy decomposition in a model-independent fashion and how these fields are related to the fundamental fields. Our key result is that one cannot always define an ``adiabatic sound speed'' as, excluding special cases such as where (\ref{eq:psisigmacondition}) is satisfied, the adiabatic field will depend on two distinct propagation speeds associated with the $\psi$ and $\omega$ fields. This has implications for the computation of quantities like the curvature power spectrum (\ref{eq:multispeedspectrum}), and may lead to interesting phenomenology. Finally, we considered several special cases such as $P(X)$ and $P(X^{\phi \phi}, X^{\chi \chi})$ Lagrangians that have been explored in previous work, recovering their results as particular limits of our work, and looked at the kinetic structure in physically-motivated scenarios, finding that supergravity does not easily provide a rich kinetic structure in which to study multi-speed inflation, but disformal couplings do naturally give rise to interesting kinetic interactions and nonlinearities.

\section*{Acknowledgements}

I would like to thank Carsten van de Bruck for our stimulating discussions during the formulation of this work. I also give my thanks to an anonymous referee for constructive and useful feedback during the peer review process. CL is supported by a STFC studentship. 

\bibliography{AdEnBib}

\end{document}